\documentclass[prb,twocolumn,showpacs,floatfix,amsmath,amssymb,superscriptaddress]{revtex4-2}
\usepackage{amsfonts}
\usepackage{stmaryrd}
\usepackage{bbm}
\usepackage{mathrsfs}
\usepackage{tipa}
\usepackage{amssymb}
\usepackage{txfonts}
\usepackage{graphicx}
\usepackage{dcolumn}
\usepackage{epstopdf}
\usepackage[colorlinks,linkcolor=blue,urlcolor=blue,citecolor=blue]{hyperref}
\usepackage{multirow}
\usepackage{subfigure}
\usepackage{url}
\usepackage{upgreek}
\usepackage[utf8]{inputenc}
\usepackage[english]{babel}
\usepackage{siunitx}

\begin{document}

\title{Quantum spin dynamics of the honeycomb magnet K$_2$Co$_2$TeO$_6$ in high magnetic fields}
% Force line breaks with \\

\author{Patrick~Pilch}
\email{patrick.pilch@tu-dortmund.de}
\affiliation{Department of Physics, TU Dortmund University, 44227 Dortmund, Germany}

\author{Laur~Peedu}
\author{Urmas~Nagel}
\author{Toomas~Rõõm}
\affiliation{National Institute of Chemical Physics and Biophysics, 12618 Tallinn, Estonia}

\author{Changqing~Zhu}
\affiliation{Department of Physics, TU Dortmund University, 44227 Dortmund, Germany}

\author{Yurii Skourski}
\affiliation{Dresden High Magnetic Field Laboratory (HLD-EMFL), Helmholtz-Zentrum Dresden–Rossendorf, D-01328 Dresden, Germany}

\author{Xianghan Xu}
\affiliation{School of Physics and Astronomy, University of Minnesota, Twin Cities, MN 55455, USA}
\affiliation{Department of Chemistry, Princeton University, New Jersey 08544, USA}
\author{Robert J. Cava}
\affiliation{Department of Chemistry, Princeton University, New Jersey 08544, USA}

\author{Zhe Wang}
\email{zhe.wang@tu-dortmund.de}
\affiliation{Department of Physics, TU Dortmund University, 44227 Dortmund, Germany}

\date{\today}

\begin{abstract}
We present terahertz spectroscopic measurements of quantum spin dynamics in the honeycomb magnet K$_2$Co$_2$TeO$_6$ as a function of temperature, polarization and in an external magnetic field applied in the honeycomb plane. Magnetic excitations are resolved below the magnetic ordering temperature of $T_\text{N} = 12$~K. In the applied magnetic field, we reveal characteristic field dependence not only for the magnetic excitations observed at zero field, but also a rich set of modes emerging in finite fields.
The observed magnetic excitations exhibit clear dependence on the terahertz polarization, and characteristic features at field-induced phase transitions consistent with our high-field magnetization data.
We cannot evidently resolve a continuumlike feature, even when the long-range magnetic order is presumably suppressed in the strong magnetic field, indicating that a Kitaev-type interaction, if existing, is subleading in this compound. 
\end{abstract}

\maketitle
\section{Introduction}
Exchange anisotropy between magnetic ions in quantum magnets often arises from a joint effect of crystal-field interaction and spin-orbit coupling \cite{AbraBlea1970}.
By tuning the crystal-field environment and/or selecting proper magnetic ions, one can realize spin interaction models beyond the isotropic Heisenberg limit in magnetic compounds to study exotic quantum phenomena.
The experimental realization of an Ising exchange anisotropy is one of the most frequently investigated problems in quantum magnetism.
For instance, in one dimension the presence of Ising anisotropy in addition to the isotropic Heisenberg exchange leads to a paradigmatic model system -- the Heisenberg-Ising spin chain, which is characterized by a spin excitation gap at zero field and undergoes a quantum phase transition in an applied transverse magnetic field \cite{Pfeuty70,Mussardo,Sachdev,Dutta,Wang2018L}.
Exotic quantum spin dynamics such as fractionalized spinon excitations and complex many-body magnon bound states can be investigated in the Heisenberg-Ising spin chain compounds (see e.g. \cite{Wang2016,Wang2018,Bera2020,Wang2024}).
In two dimensions, the so-called quantum compass model is realized by bond-dependent Ising exchange interaction \cite{Dorier2005,Jackeli2009Mott}.
In particular, on a honeycomb lattice the spin-1/2 quantum compass model, known as the Kitaev honeycomb model, was found to be exactly solvable and exhibit a quantum spin liquid state and fractionalized excitations \cite{kitaev2006anyons}.

Tremendous efforts have recently been spent to realize the Kitaev honeycomb model in quantum magnetic compounds and to investigate the intriguing physical properties (for reviews, see e.g.~\cite{wen2019experimental, Takagi2019, Broholm2020QSL,kim2021spin}).
Especially for the magnetic ions of $5d$ iridium and $4d$ ruthenium with strong spin-orbit coupling, a considerably large bond-dependent Ising anisotropy can exist in the representative compounds of Na$_2$IrO$_3$ and $\alpha$-RuCl$_3$, enabling the observation of possible quantum spin liquid states \cite{wen2019experimental, Takagi2019, Broholm2020QSL}.
Recently, the possibility to realize the Kitaev honeycomb model has also been investigated in compounds based on the $3d^7$ Co$^{2+}$ ions \cite{Liu2018Pseudospin, Sano2018Kitaev}, such as Na$_2$Co$_2$TeO$_6$ \cite{VICIU20071060,Simonet16,Bera17} and BaCo$_2$(AsO$_4)_2$ \cite{zhong2020weak,Zhang2023,Halloran23}. Although the spin-orbit coupling is reduced in comparison with the $4d$ or $5d$ elements, the existence of Ising-type anisotropy is well known for the Co$^{2+}$ ions in an octahedral crystal field with a high spin $t_{2g}^5e_g^2$ electronic configuration \cite{AbraBlea1970}.
Nonetheless, it remains an experimental problem if in the Co$^{2+}$ compounds the Ising-type anisotropy can vary from one honeycomb-lattice bond to another or is rather bond-independent.
An important signature of a quantum spin liquid state is the magnetic continuum of fractionalized excitations \cite{wen2019experimental, Takagi2019, Broholm2020QSL}. By using terahertz spectroscopy or inelastic neutron scattering, a magnetic continuum has been observed in Na$_2$Co$_2$TeO$_6$ \cite{Xiang2023Disorder,Pilch2023_Field} and BaCo$_2$(AsO$_4)_2$ \cite{Zhang2023,Halloran23} especially when the long-range magnetic order is suppressed by an applied magnetic field, similar to the observation in $\alpha$-RuCl$_3$ \cite{Wang2017alphaRuCl_FTIR,Banerjee2018}.
However, it is not a trivial task to unambiguously associate the magnetic continuum to Kitaev-type interaction in the microscopic spin Hamiltonian \cite{Halloran23,Xiang2023Disorder}, which requires a detailed spectroscopic measurement of the quantum spin dynamics. 

In this work, we investigate a recently grown honeycomb magnet K$_2$Co$_2$TeO$_6$ by performing terahertz spectroscopic measurements of quantum spin excitations. Although with a similar chemical formula as Na$_2$Co$_2$TeO$_6$ (space group $P6_322$) \cite{VICIU20071060,Simonet16,Bera17}, the K$_2$Co$_2$TeO$_6$ compound in the present study crystallizes in a different space group of $P6_3/mcm$ due to a distinct stacking type \cite{Xu2023KCTO}. The honeycomb lattice in the crystallographic \textit{ab} plane is formed by edge-sharing CoO$_6$ octahedra, as shown in the inset of Fig.~\ref{fig1}(b).
The previous measurements of magnetic susceptibility and specific heat indicate that at low temperature the magnetic ground state corresponds to an effective spin-1/2 of each Co$^{2+}$ ion, and a long-range magnetic order is formed below $T_\text{N}=12.3$~K followed by another magnetic phase transition at 4.1~K \cite{Xu2023KCTO}.
In a magnetic field applied in the honeycomb plane, three field-induced magnetic transitions were signaled by the observed anomalies in the field-dependent magnetization curve, before the spin moments are fully aligned along the field direction \cite{Xu2023KCTO}.
These interesting thermodynamic features motivate us to study the quantum spin dynamics in this compound.
By performing terahertz spectroscopic measurements as a function of temperature and in applied magnetic fields up to 17~T, we resolve the spin excitations in the low-temperature magnetically ordered phases and reveal very rich spin dynamic features characterizing the high-field induced states.

\section{Experimental details}
High-quality single crystals of K$_2$Co$_2$TeO$_6$ with an $ab$-surface were grown by a flux method, as detailed in Ref.~\cite{Xu2023KCTO}. A uniform potassium distribution was achieved by an annealing process. The samples have typical dimensions of $4 \times 3$~mm$^2$ and a thickness of about 150~$\upmu$m.
High-field magnetization measurements were performed in pulsed fields up to 30~T for various temperatures down to 1.4~K.
Terahertz absorption spectra at zero field were measured as a function of temperature by using a time-resolved terahertz spectrometer (Menlo Systems).
For the field-dependent measurements up to 17~T, the sample was installed in a cryostat equipped with a superconducting magnet and the sample temperature was maintained at 2.5~K.
The terahertz absorption scans in Voigt geometry were recorded with a Martin-Puplett interferometer coupled to a $^3$He-cooled Si-bolometer \cite{Zhang20,Amelin20,Amelin22,Pilch25}.
The direction of linear polarization of the terahertz beam was selected with a rotatable polarizer.
The terahertz polarization and the magnetic field orientation were along one of the three equivalent $a$ or one of the three $a^*$ axes [see inset of of Fig.~\ref{fig1}(b)], which are perpendicular or parallel to the Co$^{2+}$ hexagon edge, respectively.

\section{Experimental results and discussion}
\subsection{Magnetization}
\begin{figure}[t]
\centering
\includegraphics[width=1.0\linewidth]{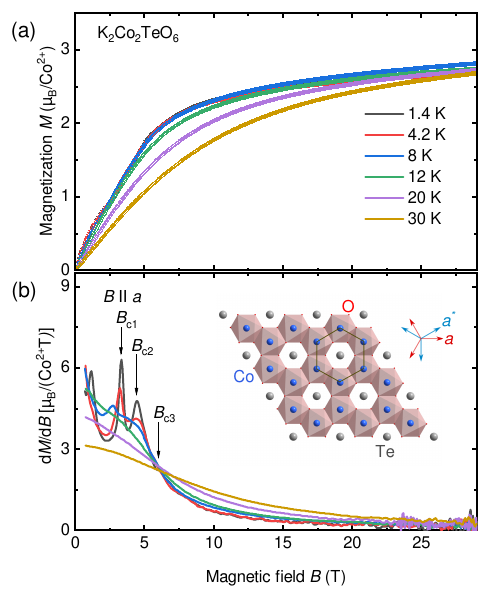}
\caption{Magnetic properties of K$_2$Co$_2$TeO$_6$. (a) High-field magnetization measured in magnetic fields up to 30~T. (b) Susceptibility calculated from (a). Two magnetic phase transitions are indicated by $B_{c1}$ and $B_{c2}$ and a third critical field is marked as $B_{c3}$. Inset: Honeycomb plane of K$_2$Co$_2$TeO$_6$ with edge-sharing CoO$_6$ octahedra. The $a$ and $a^*$ axes are perpendicular and parallel to the hexagon edges, respectively.}
\label{fig1}
\end{figure}

To show the magnetic field-induced effects, we present the high-field magnetization measured along the field direction in Fig.~\ref{fig1}(a) for various temperatures below and above the magnetic ordering temperature of about 12~K.
The corresponding magnetic susceptibility derived as field derivative of the magnetization is presented in Fig.~\ref{fig1}(b).
Here the magnetic field is applied along the crystallographic $a$ direction, i.e. $B \parallel a$. The magnetization for $B \parallel a^*$ is essentially identical as shown in Ref.~\cite{Xu2023KCTO}.
At the lowest temperature of 1.4~K the magnetization increases continuously with increasing field, and above about $B_s=18$~T it follows a paramagnetic-type linear field dependence. The corresponding magnetization value of $M_s=$~2.33~$\mathrm{\mu_B/Co^{2+}}$ is typical for a spin-1/2 ground state of Co$^{2+}$ ions in an octahedral crystal field \cite{lin2021field}.
Hence, $B_s=18$~T can be tentatively assigned as an in-plane saturation field for K$_2$Co$_2$TeO$_6$, although no anomaly in the magnetic susceptibility curve can be observed at the corresponding field.

Anomalies in the magnetic susceptibility curve are observed at lower fields.
At $B_{c1}$~=~3.3~T and $B_{c2}$~=~4.4~T two peaks are evidently seen and indicated by the arrows in Fig.~\ref{fig1}(b).
With increasing temperature these peaks broaden and merge together at 8~K, whereas above the magnetic ordering temperature of 12~K they cannot be identified anymore.
This corresponds to the temperature dependent evolution of the magnetization curves:
while at the lowest temperatures the magnetization curve exhibits kink-like features at $B_{c1}$ and $B_{c2}$, only a smooth monotonic increase of magnetization appears in the high-temperature magnetization curves above 12~K.

A third critical field $B_{c3}=6$~T, marked by an arrow in Fig.~\ref{fig1}(b), is not  a peak-like
anomaly but is rather featured as a crossing point of susceptibility  curves measured at different temperatures.
Such an isosbestic point has been observed and discussed in various contexts \cite{Greger13}, which reflects a leading dependence on one parameter, e.g. due to correlation effects \cite{Vollhardt97,Wang2014,Kovalev21}.
In the present context, the crossing point reflects a weak temperature dependence of the susceptibility at $B_{c3}$, even above $T_\text{N}$, but a relatively strong dependence on the magnetic field.
Since the peaks corresponding to $B_{c1}$ and $B_{c2}$ disappear above the three-dimensional magnetic ordering temperature, it might be natural to assume that the suppression of the long-range magnetic order by magnetic field happens at $B_{c2}$.
Consequently, the critical field of $B_{c3}$ should reflect an energy scale of the two-dimensional spin correlations, which persists even above the ordering temperature $T_\text{N}$.
As will be discussed below, we indeed observe clear contrast in quantum spin dynamics above and below $B_{c3}$.

\subsection{Quantum spin dynamics}
\begin{figure}[b]
\centering
\includegraphics[width=1.0\linewidth]{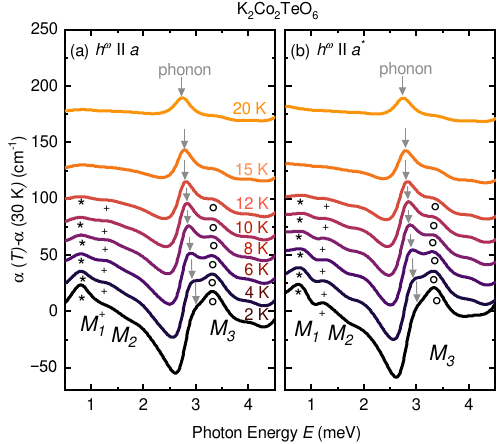}
\caption{Temperature dependence of the terahertz absorption spectra for the two polarizations (a)  $h^\omega \parallel a$ and (b) $h^\omega \parallel a^*$ (see Fig.~\ref{fig1} inset) in zero field. Three magnetic excitations are identified, $M_1$ (\textsf{*}), $M_2$(+), and $M_3$ ($\circ$). A phonon mode is marked by grey arrows. The curves are offset proportional to the temperature.}
\label{fig2}
\end{figure}

Magnetic-dipole active spin excitations can be directly observed from terahertz absorption spectra due to a linear coupling of the spin operator to the terahertz magnetic field $h^\omega$.
We show the results for various temperatures $T\leq$~20~K and zero-field in Fig.~\ref{fig2}. 
To resolve the temperature dependence of the spin excitations, we calculated the differential absorption $\alpha(T)-\alpha(30~\text{K})=-(1/d)\ln(I(T)/I(30~\text{K}))$, where $d$ is the thickness of the sample, $I(T)$ is the transmitted intensity measured at temperature $T$, and $I(30~\text{K})$ is the transmitted intensity measured at 30~K.
Above $T_\text{N}$, at 20~K, the absorption spectra for both polarizations $h^\omega\parallel a$ [Fig.~\ref{fig2}(a)] and $h^\omega\parallel a^*$ [Fig.~\ref{fig2}(b)] are characterized by a single mode peaked at 2.74~meV.
This mode can be assigned as a phonon mode, which becomes stronger and shifts to higher energy with decreasing temperature, as indicated by the arrow in Fig.~\ref{fig2}.

At the lowest temperature of 2~K which is well below $T_\text{N}$, although the phonon mode is the most dominant feature, three additional modes can be identified in both polarizations corresponding to the eigenenergies of 0.81~meV, 1.24~meV, and 3.33~meV. These three modes can be assigned as spin excitations and are denoted by $M_{1}$, $M_2$, and $M_3$ in Fig.~\ref{fig2}. 
With increasing temperatures these modes weaken and vanish above $T_\text{N}$.
No abrupt change is observed when crossing 4~K, although at this temperature another possible magnetic phase transition was suggested \cite{Xu2023KCTO}.
No spectral weight can be resolved above 4.5~meV where strong phonon absorption prevails.

\begin{figure}[tb]
\centering
\includegraphics[width=1\linewidth]{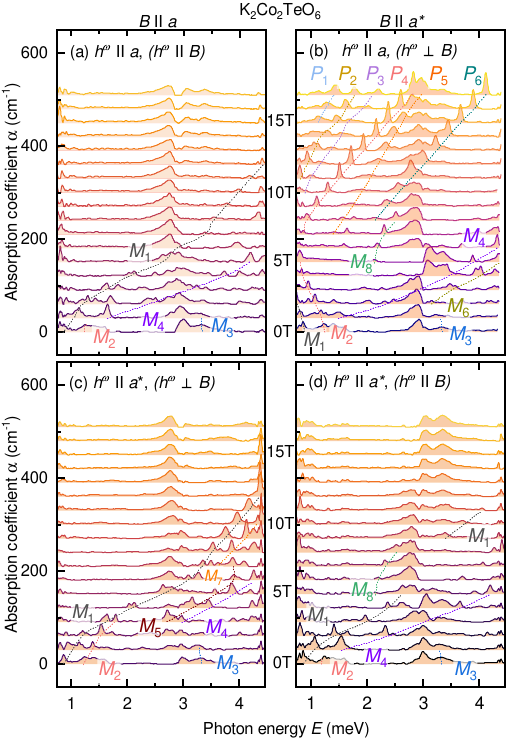}
\caption{Field-dependent terahertz absorption spectra of spin excitations measured at 2.5~K in four configurations: (a) $B \parallel a$, $h^\omega \parallel a$, (b) $B \parallel a^*$, $h^\omega \parallel a$,
(c) $B \parallel a$, $h^\omega \parallel a^*$,
and (d) $B \parallel a^*$, $h^\omega \parallel a^*$.
The spectra of higher fields with 1~T steps are shifted upwards by a constant.
The observed modes are denoted by $M_i$ or $P_i$ ($i=1,2,3,...$). The dotted lines are a guide to the eye.}
\label{fig3}
\end{figure}

To study the quantum spin dynamics in the field-induced phases, we measure terahertz absorption spectra up to 17~T in different polarizations and field orientations. Depending on whether the applied magnetic field $B$ and the terahertz magnetic field $h^\omega$ are parallel or perpendicular to the crystallographic $a$ axis, four different configurations were investigated as displayed in
the four panels of Fig.~\ref{fig3}. In these plots the field-dependent evolution of the derived absorption spectra is shown in 1~T steps.
For clarity, these spectra are offset proportionally to the applied field strength.
As a reference, the 0~T spectrum was used to calculate the differential absorption $\Delta\alpha(B)=\alpha(B)-\alpha(0~\text{T})$. $\alpha(0~\text{T})$ was recovered as the minimum of negative values of  all $\Delta\alpha(B)$ at each frequency and added to the differential absorption to obtain $\alpha(B)$.

\begin{figure*}[htb]
\centering
\includegraphics[]{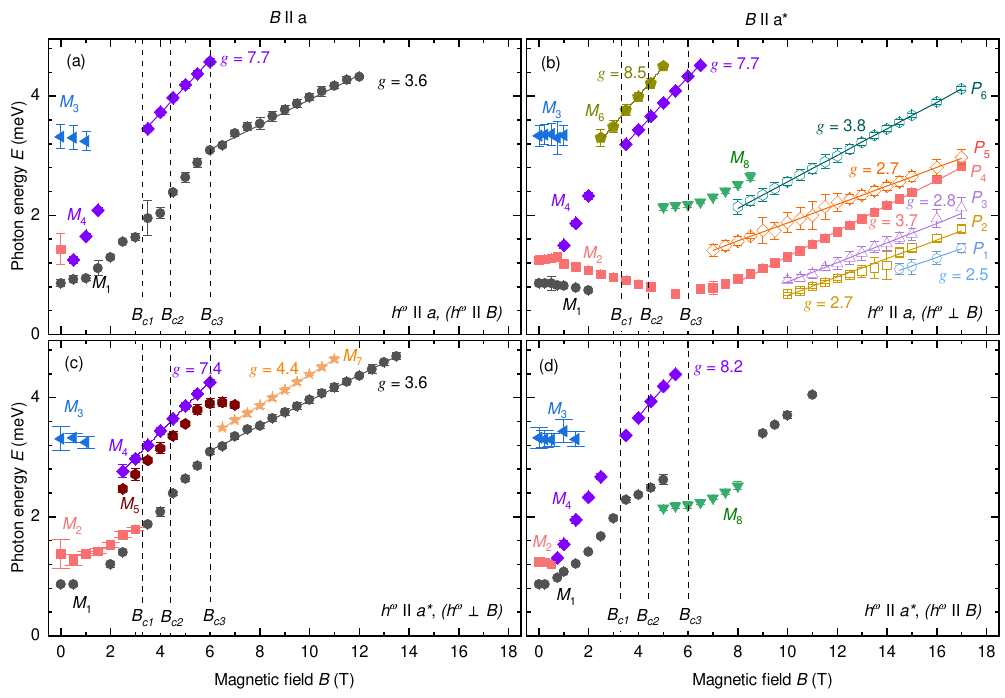}
\caption{Eigenenergies of the terahertz spin excitations in the four different configurations (a) $B\parallel a, h^\omega\parallel a$, (b) $B\parallel a^*, h^\omega\parallel a$, (c) $B\parallel a, h^\omega\parallel a^*$, and (d) $B\parallel a^*, h^\omega\parallel a^*$. The linewidths are indicated by the bars. The solid lines show a linear fit of the field dependence by a Zeeman term with the corresponding Landé $g$-factors. The three critical fields $B_{c1}$, $B_{c2}$, and $B_{c3}$ are marked by dashed lines.}
\label{fig6}
\end{figure*}

Besides the three magnetic excitation modes observed already at zero field, more modes appear at higher fields, which are very well resolved by their characteristic field dependence.
The field-dependent evolution of the observed modes is indicated by the dotted lines in Fig.~\ref{fig3}.
To provide an overview of all observed modes and their field dependence, we summarize their eigenfrequencies versus magnetic field for the four different configurations in Fig.~\ref{fig6}.
While very rich features can be directly seen from these representations, we highlight the most prominent ones as follows.

We focus first on the spectra obtained in configuration $B \parallel a^*$, $h^\omega \parallel a$ [see Fig.~\ref{fig3}(b)], which exhibits the largest number of excitation modes and different field-dependent features in comparison with spectra of the other three configurations in Fig.~\ref{fig3}(a)(c)(d).
As shown in Fig.~\ref{fig3}(b), at low fields the $M_1$ mode softens with increasing field, and out of our low-frequency spectral limit already above 2~T. 
The $M_2$ mode is split into two modes, one softening as denoted by $M_2$ and the other hardening as denoted by $M_4$.
The softening of the $M_2$ mode until 6~T is followed by an increase of its eigenenergy towards higher fields, indicating a field-induced phase transition above $B_{c3} = 6$~T [see also Fig.~\ref{fig6}(b)].
In the high-field phase beside this mode (denoted as $P_4$), five additional modes $P_i$ with $i=1,2,3,5,6$ are clearly resolvable.
These six modes cannot be observed for the other three configurations [see Fig.~\ref{fig3}(a)(c)(d)]. They exhibit a linear field dependence in the high field limit, which can be described by a Zeeman interaction term with a Landé \textit{g}-factor ranging from 2.5 to 3.8, as shown in Fig.~\ref{fig6}(b).
These \textit{g} values are typical for single magnon excitations of spin-1/2 Co$^{2+}$ ions in an octahedral crystal field \cite{AbraBlea1970}.
In contrast, the hardening of the $M_4$ mode is much more pronounced, which can be followed up to the strong phonon absorption band.
Above $B_{c1}$ this mode exhibits also a linear field dependence, corresponding to a \textit{g}-factor of 7.7 [see Fig.~\ref{fig6}(b)], which is about twice large as for the $P_6$. This strongly indicates that the $M_4$ mode could be a two-magnon excitation \cite{Wang2017alphaRuCl_FTIR,Pilch2023_Field}.
At a slightly higher energy, another mode denoted by $M_6$ [see Figs.~\ref{fig3}(b) and \ref{fig6}(b)] has an even greater \textit{g}-factor of 8.5, which might also be an excitation of two-magnon state.
An unambiguous clarification of these modes requires a detailed comparison with a microscopic spin Hamiltonian.

Although very different spin excitations emerge in the field-induced phase above $B_{c3}$ than below, no magnetic continuum can be evidently resolved in contrast to Na$_2$Co$_2$TeO$_6$ \cite{Pilch2023_Field}, BaCo$_2$(AsO$_4)_2$ \cite{Zhang2023}, and $\alpha$-RuCl$_3$ \cite{Wang2017alphaRuCl_FTIR}.
This indicates that the Kitaev-type interaction is subleading in K$_2$Co$_2$TeO$_6$, if it cannot be fully ruled out. 
The presence of the infrared phonon absorption around 3~meV reduces our resolution of the magnetic absorption in this spectral range. A measurement by inelastic neutron scattering or Raman spectroscopy could provide complementary information about the spin dynamics.
 
Without changing the external field orientation of $B \parallel a^*$, a tune of the terahertz polarization from being perpendicular to parallel to the field (i.e. $h^\omega \parallel a^* \parallel B$) modifies the spin excitation spectra [see Fig.~\ref{fig3}(d) and Fig.~\ref{fig6}(d)]. 
In the high-field phase ($B > B_{c3}$) essentially all the $P_i$ modes become inactive, together with the disappearance of the softening $M_2$ mode, which indicates a clear selection rule of these excitations.
Starting from zero field, the $M_1$ mode shifts monotonically towards higher energy, in contrast to the softening observed for $h^\omega \parallel a$.
In addition, the $M_6$ mode is not visible either in this polarization. 
For $B \parallel a^*$ at 5~T a mode at $2.14$~meV is discernible for both terahertz polarizations, as denoted by $M_8$ in Fig.~\ref{fig3}(b)(d).
In higher fields this mode shifts to higher energy into the phonon absorption band and cannot be followed above 9~T.
Since this mode cannot be found below $B_{c2}$, its field dependence probably reflects the field-induced changes at $B_{c2}$.

While keeping the configuration $h^\omega \parallel B$ but rotating the sample by 30 degrees (i.e. $h^\omega \parallel B \parallel a$), we do not see any significant changes in the spin excitation spectra when comparing Fig.~\ref{fig3}(a) with Fig.~\ref{fig3}(d).
A noticeable difference is the absence of the $M_8$ mode [Fig.~\ref{fig3}(a) and Fig.~\ref{fig6}(a)], which cannot be resolved in the corresponding spectral range.
A clear signature for the field-induced state above $B_{c3}$ is manifested by the field-dependent evolution of the $M_1$ mode.
Above $B_{c3}$, the increase of the $M_1$ eigenenergy with field is evidently reduced in comparison with that below $B_{c3}$.

Last but not least, we comment on the spin dynamics resolved in the configuration of $h^\omega \parallel a^*$, $B \parallel a$ (i.e. $h^\omega \perp B$), as presented in Fig.~\ref{fig3}(c) and Fig.~\ref{fig6}(c).
The spectra of $h^\omega \perp B$ are similar to that of $h^\omega \parallel B \parallel a$ in Fig.~\ref{fig3}(a), in a sense that $M_1$, $M_2$, $M_3$, and $M_4$ are also observed here with a similar field dependence.
In particular, the $M_1$ eigenenergy increases monotonically but the field-dependent slope reduces clearly at $B_{c3}$. The corresponding linear dependence towards higher field can be fitted by a Zeeman interaction term with a \textit{g}-factor of 3.6, comparable to that of the $P_6$ or $P_4$ mode in Fig.~\ref{fig6}(b), which indicates a field-induced reduction of many-body spin fluctuations above $B_{c3}$.
The major difference in the spectra of $h^\omega \perp B$ is the emergence of two modes above 3~T, as denoted by $M_5$ and $M_7$ [see Fig.~\ref{fig3}(c) and Fig.~\ref{fig6}(c)], which are not resolved for the other three configurations.
While the $M_7$ mode appearing above $B_{c3}$ is characteristic for the high field phase, the $M_5$ mode can be tracked only in the intermediate field range.
Another interesting feature is the behavior of the $M_2$ and $M_1$ modes below $B_{c1}$.
With increasing field the $M_2$ mode is merging into the $M_1$ mode towards $B_{c1}$ [Fig.\ref{fig3}(c)].

\section{Conclusion}
To summarize, we have performed high-field magnetization measurements and a terahertz spectroscopic study of quantum spin dynamics in the quasi-two-dimensional honeycomb magnet K$_2$Co$_2$TeO$_6$ as a function of temperature, terahertz polarization, and in an external magnetic field applied in the crystallographic honeycomb plane.
Our magnetic susceptibility curves show evidence of three critical magnetic fields of $B_{c1}=3.3$~T, $B_{c2}=4.4$~T, and $B_{c3}=6$~T.
While the former two are identified by peaks in the low-temperature susceptibility curves below $T_\text{N} = 12$~K, the last one corresponds to an isosbestic point where the curves of different temperatures even above $T_\text{N}$ intersect.
This indicates a field-induced suppression of the long-range magnetic order at $B_{c2}$, whereas a saturation of the magnetization is reached above about 
$B_{s}=18$~T.

At zero field, the temperature-dependent terahertz spectra reveal characteristic magnetic excitations for the magnetically ordered phase below $T_\text{N} = 12$~K.
In finite magnetic fields, more spin excitation modes are observed and followed until 17~T, crossing the critical fields of $B_{c1}$,  $B_{c2}$, and $B_{c3}$.  
The field-dependent evolution of the magnetic excitations varies also with the terahertz polarization and field orientation, and exhibits features evidencing the field-induced phase transitions.
At the field-induced phase transitions, we cannot observe an evident feature of a magnetic continuum, which suggests that this compound does not have a significant Kitaev-type interaction.
Our results call for complementary measurements of the spin dynamics, e.g. by inelastic neutron scattering or Raman spectroscopy, and also model simulation of a spin interaction Hamiltonian to identify the relevance of anisotropic Heisenberg interactions and possible Kitaev-type interaction.\\

\begin{acknowledgments}
We acknowledge support by the European Research Council (ERC) under the Horizon 2020 research and innovation programme, Grant Agreement No. 950560 (DynaQuanta), by the Estonian Ministry of Education, personal research funding PRG736, and by the European Regional Development Fund project TK134. This work was supported by the HLD at HZDR, member of the European Magnetic Field Laboratory (EMFL).
\end{acknowledgments}

\bibliography{KCTO_bib}% Produces the bibliography vi BibTeX.

\begin{thebibliography}{39}%
\makeatletter
\providecommand \@ifxundefined [1]{%
 \@ifx{#1\undefined}
}%
\providecommand \@ifnum [1]{%
 \ifnum #1\expandafter \@firstoftwo
 \else \expandafter \@secondoftwo
 \fi
}%
\providecommand \@ifx [1]{%
 \ifx #1\expandafter \@firstoftwo
 \else \expandafter \@secondoftwo
 \fi
}%
\providecommand \natexlab [1]{#1}%
\providecommand \enquote  [1]{``#1''}%
\providecommand \bibnamefont  [1]{#1}%
\providecommand \bibfnamefont [1]{#1}%
\providecommand \citenamefont [1]{#1}%
\providecommand \href@noop [0]{\@secondoftwo}%
\providecommand \href [0]{\begingroup \@sanitize@url \@href}%
\providecommand \@href[1]{\@@startlink{#1}\@@href}%
\providecommand \@@href[1]{\endgroup#1\@@endlink}%
\providecommand \@sanitize@url [0]{\catcode `\\12\catcode `\$12\catcode
  `\&12\catcode `\#12\catcode `\^12\catcode `\_12\catcode `\%12\relax}%
\providecommand \@@startlink[1]{}%
\providecommand \@@endlink[0]{}%
\providecommand \url  [0]{\begingroup\@sanitize@url \@url }%
\providecommand \@url [1]{\endgroup\@href {#1}{\urlprefix }}%
\providecommand \urlprefix  [0]{URL }%
\providecommand \Eprint [0]{\href }%
\providecommand \doibase [0]{https://doi.org/}%
\providecommand \selectlanguage [0]{\@gobble}%
\providecommand \bibinfo  [0]{\@secondoftwo}%
\providecommand \bibfield  [0]{\@secondoftwo}%
\providecommand \translation [1]{[#1]}%
\providecommand \BibitemOpen [0]{}%
\providecommand \bibitemStop [0]{}%
\providecommand \bibitemNoStop [0]{.\EOS\space}%
\providecommand \EOS [0]{\spacefactor3000\relax}%
\providecommand \BibitemShut  [1]{\csname bibitem#1\endcsname}%
\let\auto@bib@innerbib\@empty
%</preamble>
\bibitem [{\citenamefont {Abragam}\ and\ \citenamefont
  {Bleaney}(1970)}]{AbraBlea1970}%
  \BibitemOpen
  \bibfield  {author} {\bibinfo {author} {\bibfnamefont {A.}~\bibnamefont
  {Abragam}}\ and\ \bibinfo {author} {\bibfnamefont {B.}~\bibnamefont
  {Bleaney}},\ }\href@noop {} {\emph {\bibinfo {title} {Electron Paramagnetic
  Resonance of Transition Ions}}}\ (\bibinfo  {publisher} {Clarendon Press},\
  \bibinfo {address} {Oxford},\ \bibinfo {year} {1970})\BibitemShut {NoStop}%
\bibitem [{\citenamefont {Pfeuty}(1970)}]{Pfeuty70}%
  \BibitemOpen
  \bibfield  {author} {\bibinfo {author} {\bibfnamefont {P.}~\bibnamefont
  {Pfeuty}},\ }\bibfield  {title} {\bibinfo {title} {The one-dimensional
  {I}sing model with a transverse field},\ }\href
  {https://doi.org/10.1016/0003-4916(70)90270-8} {\bibfield  {journal}
  {\bibinfo  {journal} {Ann. Phys. (N.Y.)}\ }\textbf {\bibinfo {volume} {57}},\
  \bibinfo {pages} {79} (\bibinfo {year} {1970})}\BibitemShut {NoStop}%
\bibitem [{\citenamefont {Mussardo}(2010)}]{Mussardo}%
  \BibitemOpen
  \bibfield  {author} {\bibinfo {author} {\bibfnamefont {G.}~\bibnamefont
  {Mussardo}},\ }\href@noop {} {\emph {\bibinfo {title} {Statistical Field
  Theory}}}\ (\bibinfo  {publisher} {Oxford University Press},\ \bibinfo
  {address} {New York},\ \bibinfo {year} {2010})\BibitemShut {NoStop}%
\bibitem [{\citenamefont {Sachdev}(2011)}]{Sachdev}%
  \BibitemOpen
  \bibfield  {author} {\bibinfo {author} {\bibfnamefont {S.}~\bibnamefont
  {Sachdev}},\ }\href@noop {} {\emph {\bibinfo {title} {Quantum Phase
  Transitions}}},\ \bibinfo {edition} {2nd}\ ed.\ (\bibinfo  {publisher}
  {Cambridge University Press},\ \bibinfo {address} {New York},\ \bibinfo
  {year} {2011})\BibitemShut {NoStop}%
\bibitem [{\citenamefont {Dutta}\ \emph {et~al.}(2015)\citenamefont {Dutta},
  \citenamefont {Aeppli}, \citenamefont {Chakrabarti}, \citenamefont
  {Divakaran}, \citenamefont {Rosenbaum},\ and\ \citenamefont {Sen}}]{Dutta}%
  \BibitemOpen
  \bibfield  {author} {\bibinfo {author} {\bibfnamefont {A.}~\bibnamefont
  {Dutta}}, \bibinfo {author} {\bibfnamefont {G.}~\bibnamefont {Aeppli}},
  \bibinfo {author} {\bibfnamefont {B.~K.}\ \bibnamefont {Chakrabarti}},
  \bibinfo {author} {\bibfnamefont {U.}~\bibnamefont {Divakaran}}, \bibinfo
  {author} {\bibfnamefont {T.~F.}\ \bibnamefont {Rosenbaum}},\ and\ \bibinfo
  {author} {\bibfnamefont {D.}~\bibnamefont {Sen}},\ }\href@noop {} {\emph
  {\bibinfo {title} {Quantum Phase Transitions in Transverse Field Spin Models:
  From Statistical Physics to Quantum Information}}}\ (\bibinfo  {publisher}
  {Cambridge University Press},\ \bibinfo {address} {Cambridge},\ \bibinfo
  {year} {2015})\BibitemShut {NoStop}%
\bibitem [{\citenamefont {Wang}\ \emph
  {et~al.}(2018{\natexlab{a}})\citenamefont {Wang}, \citenamefont {Lorenz},
  \citenamefont {Gorbunov}, \citenamefont {Cong}, \citenamefont {Kohama},
  \citenamefont {Niesen}, \citenamefont {Breunig}, \citenamefont {Engelmayer},
  \citenamefont {Herman}, \citenamefont {Wu}, \citenamefont {Kindo},
  \citenamefont {Wosnitza}, \citenamefont {Zherlitsyn},\ and\ \citenamefont
  {Loidl}}]{Wang2018L}%
  \BibitemOpen
  \bibfield  {author} {\bibinfo {author} {\bibfnamefont {Z.}~\bibnamefont
  {Wang}}, \bibinfo {author} {\bibfnamefont {T.}~\bibnamefont {Lorenz}},
  \bibinfo {author} {\bibfnamefont {D.~I.}\ \bibnamefont {Gorbunov}}, \bibinfo
  {author} {\bibfnamefont {P.~T.}\ \bibnamefont {Cong}}, \bibinfo {author}
  {\bibfnamefont {Y.}~\bibnamefont {Kohama}}, \bibinfo {author} {\bibfnamefont
  {S.}~\bibnamefont {Niesen}}, \bibinfo {author} {\bibfnamefont
  {O.}~\bibnamefont {Breunig}}, \bibinfo {author} {\bibfnamefont
  {J.}~\bibnamefont {Engelmayer}}, \bibinfo {author} {\bibfnamefont
  {A.}~\bibnamefont {Herman}}, \bibinfo {author} {\bibfnamefont
  {J.}~\bibnamefont {Wu}}, \bibinfo {author} {\bibfnamefont {K.}~\bibnamefont
  {Kindo}}, \bibinfo {author} {\bibfnamefont {J.}~\bibnamefont {Wosnitza}},
  \bibinfo {author} {\bibfnamefont {S.}~\bibnamefont {Zherlitsyn}},\ and\
  \bibinfo {author} {\bibfnamefont {A.}~\bibnamefont {Loidl}},\ }\bibfield
  {title} {\bibinfo {title} {Quantum criticality of an {I}sing-like spin-$1/2$
  antiferromagnetic chain in a transverse magnetic field},\ }\href
  {https://doi.org/10.1103/PhysRevLett.120.207205} {\bibfield  {journal}
  {\bibinfo  {journal} {Phys. Rev. Lett.}\ }\textbf {\bibinfo {volume} {120}},\
  \bibinfo {pages} {207205} (\bibinfo {year} {2018}{\natexlab{a}})}\BibitemShut
  {NoStop}%
\bibitem [{\citenamefont {Wang}\ \emph {et~al.}(2016)\citenamefont {Wang},
  \citenamefont {Wu}, \citenamefont {Xu}, \citenamefont {Yang}, \citenamefont
  {Wu}, \citenamefont {Bera}, \citenamefont {Islam}, \citenamefont {Lake},
  \citenamefont {Kamenskyi}, \citenamefont {Gogoi}, \citenamefont {Engelkamp},
  \citenamefont {Wang}, \citenamefont {Deisenhofer},\ and\ \citenamefont
  {Loidl}}]{Wang2016}%
  \BibitemOpen
  \bibfield  {author} {\bibinfo {author} {\bibfnamefont {Z.}~\bibnamefont
  {Wang}}, \bibinfo {author} {\bibfnamefont {J.}~\bibnamefont {Wu}}, \bibinfo
  {author} {\bibfnamefont {S.}~\bibnamefont {Xu}}, \bibinfo {author}
  {\bibfnamefont {W.}~\bibnamefont {Yang}}, \bibinfo {author} {\bibfnamefont
  {C.}~\bibnamefont {Wu}}, \bibinfo {author} {\bibfnamefont {A.~K.}\
  \bibnamefont {Bera}}, \bibinfo {author} {\bibfnamefont {A.~T. M.~N.}\
  \bibnamefont {Islam}}, \bibinfo {author} {\bibfnamefont {B.}~\bibnamefont
  {Lake}}, \bibinfo {author} {\bibfnamefont {D.}~\bibnamefont {Kamenskyi}},
  \bibinfo {author} {\bibfnamefont {P.}~\bibnamefont {Gogoi}}, \bibinfo
  {author} {\bibfnamefont {H.}~\bibnamefont {Engelkamp}}, \bibinfo {author}
  {\bibfnamefont {N.}~\bibnamefont {Wang}}, \bibinfo {author} {\bibfnamefont
  {J.}~\bibnamefont {Deisenhofer}},\ and\ \bibinfo {author} {\bibfnamefont
  {A.}~\bibnamefont {Loidl}},\ }\bibfield  {title} {\bibinfo {title} {From
  confined spinons to emergent fermions: Observation of elementary magnetic
  excitations in a transverse-field {I}sing chain},\ }\href
  {https://doi.org/10.1103/PhysRevB.94.125130} {\bibfield  {journal} {\bibinfo
  {journal} {Phys. Rev. B}\ }\textbf {\bibinfo {volume} {94}},\ \bibinfo
  {pages} {125130} (\bibinfo {year} {2016})}\BibitemShut {NoStop}%
\bibitem [{\citenamefont {Wang}\ \emph
  {et~al.}(2018{\natexlab{b}})\citenamefont {Wang}, \citenamefont {Wu},
  \citenamefont {Yang}, \citenamefont {Bera}, \citenamefont {Kamenskyi},
  \citenamefont {Islam}, \citenamefont {Xu}, \citenamefont {Law}, \citenamefont
  {Lake}, \citenamefont {Wu},\ and\ \citenamefont {Loidl}}]{Wang2018}%
  \BibitemOpen
  \bibfield  {author} {\bibinfo {author} {\bibfnamefont {Z.}~\bibnamefont
  {Wang}}, \bibinfo {author} {\bibfnamefont {J.}~\bibnamefont {Wu}}, \bibinfo
  {author} {\bibfnamefont {W.}~\bibnamefont {Yang}}, \bibinfo {author}
  {\bibfnamefont {A.~K.}\ \bibnamefont {Bera}}, \bibinfo {author}
  {\bibfnamefont {D.}~\bibnamefont {Kamenskyi}}, \bibinfo {author}
  {\bibfnamefont {A.~T. M.~N.}\ \bibnamefont {Islam}}, \bibinfo {author}
  {\bibfnamefont {S.}~\bibnamefont {Xu}}, \bibinfo {author} {\bibfnamefont
  {J.~M.}\ \bibnamefont {Law}}, \bibinfo {author} {\bibfnamefont
  {B.}~\bibnamefont {Lake}}, \bibinfo {author} {\bibfnamefont {C.}~\bibnamefont
  {Wu}},\ and\ \bibinfo {author} {\bibfnamefont {A.}~\bibnamefont {Loidl}},\
  }\bibfield  {title} {\bibinfo {title} {Experimental observation of {B}ethe
  strings},\ }\href {https://doi.org/10.1038/nature25466} {\bibfield  {journal}
  {\bibinfo  {journal} {Nature}\ }\textbf {\bibinfo {volume} {554}},\ \bibinfo
  {pages} {219} (\bibinfo {year} {2018}{\natexlab{b}})}\BibitemShut {NoStop}%
\bibitem [{\citenamefont {Bera}\ \emph {et~al.}(2020)\citenamefont {Bera},
  \citenamefont {Wu}, \citenamefont {Yang}, \citenamefont {Bewley},
  \citenamefont {Boehm}, \citenamefont {Xu}, \citenamefont {Bartkowiak},
  \citenamefont {Prokhnenko}, \citenamefont {Klemke}, \citenamefont {Islam},
  \citenamefont {Law}, \citenamefont {Wang},\ and\ \citenamefont
  {Lake}}]{Bera2020}%
  \BibitemOpen
  \bibfield  {author} {\bibinfo {author} {\bibfnamefont {A.~K.}\ \bibnamefont
  {Bera}}, \bibinfo {author} {\bibfnamefont {J.}~\bibnamefont {Wu}}, \bibinfo
  {author} {\bibfnamefont {W.}~\bibnamefont {Yang}}, \bibinfo {author}
  {\bibfnamefont {R.}~\bibnamefont {Bewley}}, \bibinfo {author} {\bibfnamefont
  {M.}~\bibnamefont {Boehm}}, \bibinfo {author} {\bibfnamefont
  {J.}~\bibnamefont {Xu}}, \bibinfo {author} {\bibfnamefont {M.}~\bibnamefont
  {Bartkowiak}}, \bibinfo {author} {\bibfnamefont {O.}~\bibnamefont
  {Prokhnenko}}, \bibinfo {author} {\bibfnamefont {B.}~\bibnamefont {Klemke}},
  \bibinfo {author} {\bibfnamefont {A.~T. M.~N.}\ \bibnamefont {Islam}},
  \bibinfo {author} {\bibfnamefont {J.~M.}\ \bibnamefont {Law}}, \bibinfo
  {author} {\bibfnamefont {Z.}~\bibnamefont {Wang}},\ and\ \bibinfo {author}
  {\bibfnamefont {B.}~\bibnamefont {Lake}},\ }\bibfield  {title} {\bibinfo
  {title} {Dispersions of many-body {B}ethe strings},\ }\href
  {https://doi.org/10.1038/s41567-020-0835-7} {\bibfield  {journal} {\bibinfo
  {journal} {Nat. Phys.}\ }\textbf {\bibinfo {volume} {16}},\ \bibinfo {pages}
  {625} (\bibinfo {year} {2020})}\BibitemShut {NoStop}%
\bibitem [{\citenamefont {Wang}\ \emph {et~al.}(2024)\citenamefont {Wang},
  \citenamefont {Halati}, \citenamefont {Bernier}, \citenamefont {Ponomaryov},
  \citenamefont {Gorbunov}, \citenamefont {Niesen}, \citenamefont {Breunig},
  \citenamefont {Klopf}, \citenamefont {Zvyagin}, \citenamefont {Lorenz},
  \citenamefont {Loidl},\ and\ \citenamefont {Kollath}}]{Wang2024}%
  \BibitemOpen
  \bibfield  {author} {\bibinfo {author} {\bibfnamefont {Z.}~\bibnamefont
  {Wang}}, \bibinfo {author} {\bibfnamefont {C.-M.}\ \bibnamefont {Halati}},
  \bibinfo {author} {\bibfnamefont {J.-S.}\ \bibnamefont {Bernier}}, \bibinfo
  {author} {\bibfnamefont {A.}~\bibnamefont {Ponomaryov}}, \bibinfo {author}
  {\bibfnamefont {D.~I.}\ \bibnamefont {Gorbunov}}, \bibinfo {author}
  {\bibfnamefont {S.}~\bibnamefont {Niesen}}, \bibinfo {author} {\bibfnamefont
  {O.}~\bibnamefont {Breunig}}, \bibinfo {author} {\bibfnamefont {J.~M.}\
  \bibnamefont {Klopf}}, \bibinfo {author} {\bibfnamefont {S.}~\bibnamefont
  {Zvyagin}}, \bibinfo {author} {\bibfnamefont {T.}~\bibnamefont {Lorenz}},
  \bibinfo {author} {\bibfnamefont {A.}~\bibnamefont {Loidl}},\ and\ \bibinfo
  {author} {\bibfnamefont {C.}~\bibnamefont {Kollath}},\ }\bibfield  {title}
  {\bibinfo {title} {Experimental observation of repulsively bound magnons},\
  }\href {https://doi.org/10.1038/s41586-024-07599-3} {\bibfield  {journal}
  {\bibinfo  {journal} {Nature}\ }\textbf {\bibinfo {volume} {631}},\ \bibinfo
  {pages} {760} (\bibinfo {year} {2024})}\BibitemShut {NoStop}%
\bibitem [{\citenamefont {Dorier}\ \emph {et~al.}(2005)\citenamefont {Dorier},
  \citenamefont {Becca},\ and\ \citenamefont {Mila}}]{Dorier2005}%
  \BibitemOpen
  \bibfield  {author} {\bibinfo {author} {\bibfnamefont {J.}~\bibnamefont
  {Dorier}}, \bibinfo {author} {\bibfnamefont {F.}~\bibnamefont {Becca}},\ and\
  \bibinfo {author} {\bibfnamefont {F.}~\bibnamefont {Mila}},\ }\bibfield
  {title} {\bibinfo {title} {Quantum compass model on the square lattice},\
  }\href {https://doi.org/10.1103/PhysRevB.72.024448} {\bibfield  {journal}
  {\bibinfo  {journal} {Phys. Rev. B}\ }\textbf {\bibinfo {volume} {72}},\
  \bibinfo {pages} {024448} (\bibinfo {year} {2005})}\BibitemShut {NoStop}%
\bibitem [{\citenamefont {Jackeli}\ and\ \citenamefont
  {Khaliullin}(2009)}]{Jackeli2009Mott}%
  \BibitemOpen
  \bibfield  {author} {\bibinfo {author} {\bibfnamefont {G.}~\bibnamefont
  {Jackeli}}\ and\ \bibinfo {author} {\bibfnamefont {G.}~\bibnamefont
  {Khaliullin}},\ }\bibfield  {title} {\bibinfo {title} {Mott insulators in the
  strong spin-orbit coupling limit: From {H}eisenberg to a quantum compass and
  {K}itaev models},\ }\href {https://doi.org/10.1103/PhysRevLett.102.017205}
  {\bibfield  {journal} {\bibinfo  {journal} {Phys. Rev. Lett.}\ }\textbf
  {\bibinfo {volume} {102}},\ \bibinfo {pages} {017205} (\bibinfo {year}
  {2009})}\BibitemShut {NoStop}%
\bibitem [{\citenamefont {Kitaev}(2006)}]{kitaev2006anyons}%
  \BibitemOpen
  \bibfield  {author} {\bibinfo {author} {\bibfnamefont {A.}~\bibnamefont
  {Kitaev}},\ }\bibfield  {title} {\bibinfo {title} {Anyons in an exactly
  solved model and beyond},\ }\href {https://doi.org/10.1016/j.aop.2005.10.005}
  {\bibfield  {journal} {\bibinfo  {journal} {Ann. Phys.}\ }\textbf {\bibinfo
  {volume} {321}},\ \bibinfo {pages} {2} (\bibinfo {year} {2006})}\BibitemShut
  {NoStop}%
\bibitem [{\citenamefont {Wen}\ \emph {et~al.}(2019)\citenamefont {Wen},
  \citenamefont {Yu}, \citenamefont {Li}, \citenamefont {Yu},\ and\
  \citenamefont {Li}}]{wen2019experimental}%
  \BibitemOpen
  \bibfield  {author} {\bibinfo {author} {\bibfnamefont {J.}~\bibnamefont
  {Wen}}, \bibinfo {author} {\bibfnamefont {S.-L.}\ \bibnamefont {Yu}},
  \bibinfo {author} {\bibfnamefont {S.}~\bibnamefont {Li}}, \bibinfo {author}
  {\bibfnamefont {W.}~\bibnamefont {Yu}},\ and\ \bibinfo {author}
  {\bibfnamefont {J.-X.}\ \bibnamefont {Li}},\ }\bibfield  {title} {\bibinfo
  {title} {Experimental identification of quantum spin liquids},\ }\href
  {https://doi.org/10.1038/s41535-019-0151-6} {\bibfield  {journal} {\bibinfo
  {journal} {npj Quantum Mater}\ }\textbf {\bibinfo {volume} {4}},\ \bibinfo
  {pages} {12} (\bibinfo {year} {2019})}\BibitemShut {NoStop}%
\bibitem [{\citenamefont {Takagi}\ \emph {et~al.}(2019)\citenamefont {Takagi},
  \citenamefont {Takayama}, \citenamefont {Jackeli}, \citenamefont
  {Khaliullin},\ and\ \citenamefont {Nagler}}]{Takagi2019}%
  \BibitemOpen
  \bibfield  {author} {\bibinfo {author} {\bibfnamefont {H.}~\bibnamefont
  {Takagi}}, \bibinfo {author} {\bibfnamefont {T.}~\bibnamefont {Takayama}},
  \bibinfo {author} {\bibfnamefont {G.}~\bibnamefont {Jackeli}}, \bibinfo
  {author} {\bibfnamefont {G.}~\bibnamefont {Khaliullin}},\ and\ \bibinfo
  {author} {\bibfnamefont {S.~E.}\ \bibnamefont {Nagler}},\ }\bibfield  {title}
  {\bibinfo {title} {Concept and realization of {K}itaev quantum spin
  liquids},\ }\href {https://doi.org/10.1038/s42254-019-0038-2} {\bibfield
  {journal} {\bibinfo  {journal} {Nat. Rev. Phys.}\ }\textbf {\bibinfo {volume}
  {1}},\ \bibinfo {pages} {264} (\bibinfo {year} {2019})}\BibitemShut {NoStop}%
\bibitem [{\citenamefont {Broholm}\ \emph {et~al.}(2020)\citenamefont
  {Broholm}, \citenamefont {Cava}, \citenamefont {Kivelson}, \citenamefont
  {Nocera}, \citenamefont {Norman},\ and\ \citenamefont
  {Senthil}}]{Broholm2020QSL}%
  \BibitemOpen
  \bibfield  {author} {\bibinfo {author} {\bibfnamefont {C.}~\bibnamefont
  {Broholm}}, \bibinfo {author} {\bibfnamefont {R.~J.}\ \bibnamefont {Cava}},
  \bibinfo {author} {\bibfnamefont {S.~A.}\ \bibnamefont {Kivelson}}, \bibinfo
  {author} {\bibfnamefont {D.~G.}\ \bibnamefont {Nocera}}, \bibinfo {author}
  {\bibfnamefont {M.~R.}\ \bibnamefont {Norman}},\ and\ \bibinfo {author}
  {\bibfnamefont {T.}~\bibnamefont {Senthil}},\ }\bibfield  {title} {\bibinfo
  {title} {Quantum spin liquids},\ }\href
  {https://doi.org/10.1126/science.aay0668} {\bibfield  {journal} {\bibinfo
  {journal} {Science}\ }\textbf {\bibinfo {volume} {367}},\ \bibinfo {pages}
  {eaay0668} (\bibinfo {year} {2020})}\BibitemShut {NoStop}%
\bibitem [{\citenamefont {Kim}\ \emph {et~al.}(2021)\citenamefont {Kim},
  \citenamefont {Kim},\ and\ \citenamefont {Park}}]{kim2021spin}%
  \BibitemOpen
  \bibfield  {author} {\bibinfo {author} {\bibfnamefont {C.}~\bibnamefont
  {Kim}}, \bibinfo {author} {\bibfnamefont {H.-S.}\ \bibnamefont {Kim}},\ and\
  \bibinfo {author} {\bibfnamefont {J.-G.}\ \bibnamefont {Park}},\ }\bibfield
  {title} {\bibinfo {title} {Spin-orbital entangled state and realization of
  {K}itaev physics in 3$d$ cobalt compounds: a progress report},\ }\href
  {https://doi.org/10.1088/1361-648X/ac2d5d} {\bibfield  {journal} {\bibinfo
  {journal} {J. Phys.: Condens. Matter.}\ }\textbf {\bibinfo {volume} {34}},\
  \bibinfo {pages} {023001} (\bibinfo {year} {2021})}\BibitemShut {NoStop}%
\bibitem [{\citenamefont {Liu}\ and\ \citenamefont
  {Khaliullin}(2018)}]{Liu2018Pseudospin}%
  \BibitemOpen
  \bibfield  {author} {\bibinfo {author} {\bibfnamefont {H.}~\bibnamefont
  {Liu}}\ and\ \bibinfo {author} {\bibfnamefont {G.}~\bibnamefont
  {Khaliullin}},\ }\bibfield  {title} {\bibinfo {title} {Pseudospin exchange
  interactions in ${d}^{7}$ cobalt compounds: Possible realization of the
  {K}itaev model},\ }\href {https://doi.org/10.1103/PhysRevB.97.014407}
  {\bibfield  {journal} {\bibinfo  {journal} {Phys. Rev. B}\ }\textbf {\bibinfo
  {volume} {97}},\ \bibinfo {pages} {014407} (\bibinfo {year}
  {2018})}\BibitemShut {NoStop}%
\bibitem [{\citenamefont {Sano}\ \emph {et~al.}(2018)\citenamefont {Sano},
  \citenamefont {Kato},\ and\ \citenamefont {Motome}}]{Sano2018Kitaev}%
  \BibitemOpen
  \bibfield  {author} {\bibinfo {author} {\bibfnamefont {R.}~\bibnamefont
  {Sano}}, \bibinfo {author} {\bibfnamefont {Y.}~\bibnamefont {Kato}},\ and\
  \bibinfo {author} {\bibfnamefont {Y.}~\bibnamefont {Motome}},\ }\bibfield
  {title} {\bibinfo {title} {Kitaev-{H}eisenberg {H}amiltonian for high-spin
  ${d}^{7}$ {M}ott insulators},\ }\href
  {https://doi.org/10.1103/PhysRevB.97.014408} {\bibfield  {journal} {\bibinfo
  {journal} {Phys. Rev. B}\ }\textbf {\bibinfo {volume} {97}},\ \bibinfo
  {pages} {014408} (\bibinfo {year} {2018})}\BibitemShut {NoStop}%
\bibitem [{\citenamefont {Viciu}\ \emph {et~al.}(2007)\citenamefont {Viciu},
  \citenamefont {Huang}, \citenamefont {Morosan}, \citenamefont {Zandbergen},
  \citenamefont {Greenbaum}, \citenamefont {McQueen},\ and\ \citenamefont
  {Cava}}]{VICIU20071060}%
  \BibitemOpen
  \bibfield  {author} {\bibinfo {author} {\bibfnamefont {L.}~\bibnamefont
  {Viciu}}, \bibinfo {author} {\bibfnamefont {Q.}~\bibnamefont {Huang}},
  \bibinfo {author} {\bibfnamefont {E.}~\bibnamefont {Morosan}}, \bibinfo
  {author} {\bibfnamefont {H.}~\bibnamefont {Zandbergen}}, \bibinfo {author}
  {\bibfnamefont {N.}~\bibnamefont {Greenbaum}}, \bibinfo {author}
  {\bibfnamefont {T.}~\bibnamefont {McQueen}},\ and\ \bibinfo {author}
  {\bibfnamefont {R.}~\bibnamefont {Cava}},\ }\bibfield  {title} {\bibinfo
  {title} {Structure and basic magnetic properties of the honeycomb lattice
  compounds {N}a$_2${C}o$_2${T}e{O}$_6$ and {N}a$_3${C}o$_2${S}b{O}$_6$},\
  }\href {https://doi.org/https://doi.org/10.1016/j.jssc.2007.01.002}
  {\bibfield  {journal} {\bibinfo  {journal} {J. Solid State Chem.}\ }\textbf
  {\bibinfo {volume} {180}},\ \bibinfo {pages} {1060} (\bibinfo {year}
  {2007})}\BibitemShut {NoStop}%
\bibitem [{\citenamefont {Lefran\ifmmode~\mbox{\c{c}}\else \c{c}\fi{}ois}\
  \emph {et~al.}(2016)\citenamefont {Lefran\ifmmode~\mbox{\c{c}}\else
  \c{c}\fi{}ois}, \citenamefont {Songvilay}, \citenamefont {Robert},
  \citenamefont {Nataf}, \citenamefont {Jordan}, \citenamefont {Chaix},
  \citenamefont {Colin}, \citenamefont {Lejay}, \citenamefont {Hadj-Azzem},
  \citenamefont {Ballou},\ and\ \citenamefont {Simonet}}]{Simonet16}%
  \BibitemOpen
  \bibfield  {author} {\bibinfo {author} {\bibfnamefont {E.}~\bibnamefont
  {Lefran\ifmmode~\mbox{\c{c}}\else \c{c}\fi{}ois}}, \bibinfo {author}
  {\bibfnamefont {M.}~\bibnamefont {Songvilay}}, \bibinfo {author}
  {\bibfnamefont {J.}~\bibnamefont {Robert}}, \bibinfo {author} {\bibfnamefont
  {G.}~\bibnamefont {Nataf}}, \bibinfo {author} {\bibfnamefont
  {E.}~\bibnamefont {Jordan}}, \bibinfo {author} {\bibfnamefont
  {L.}~\bibnamefont {Chaix}}, \bibinfo {author} {\bibfnamefont {C.~V.}\
  \bibnamefont {Colin}}, \bibinfo {author} {\bibfnamefont {P.}~\bibnamefont
  {Lejay}}, \bibinfo {author} {\bibfnamefont {A.}~\bibnamefont {Hadj-Azzem}},
  \bibinfo {author} {\bibfnamefont {R.}~\bibnamefont {Ballou}},\ and\ \bibinfo
  {author} {\bibfnamefont {V.}~\bibnamefont {Simonet}},\ }\bibfield  {title}
  {\bibinfo {title} {Magnetic properties of the honeycomb oxide
  {N}a$_{2}${C}o$_{2}${T}e{O}$_{6}$},\ }\href
  {https://doi.org/10.1103/PhysRevB.94.214416} {\bibfield  {journal} {\bibinfo
  {journal} {Phys. Rev. B}\ }\textbf {\bibinfo {volume} {94}},\ \bibinfo
  {pages} {214416} (\bibinfo {year} {2016})}\BibitemShut {NoStop}%
\bibitem [{\citenamefont {Bera}\ \emph {et~al.}(2017)\citenamefont {Bera},
  \citenamefont {Yusuf}, \citenamefont {Kumar},\ and\ \citenamefont
  {Ritter}}]{Bera17}%
  \BibitemOpen
  \bibfield  {author} {\bibinfo {author} {\bibfnamefont {A.~K.}\ \bibnamefont
  {Bera}}, \bibinfo {author} {\bibfnamefont {S.~M.}\ \bibnamefont {Yusuf}},
  \bibinfo {author} {\bibfnamefont {A.}~\bibnamefont {Kumar}},\ and\ \bibinfo
  {author} {\bibfnamefont {C.}~\bibnamefont {Ritter}},\ }\bibfield  {title}
  {\bibinfo {title} {Zigzag antiferromagnetic ground state with anisotropic
  correlation lengths in the quasi-two-dimensional honeycomb lattice compound
  $\mathrm{N}{\mathrm{a}}_{2}\mathrm{C}{\mathrm{o}}_{2}\mathrm{Te}{\mathrm{o}}_{6}$},\
  }\href {https://doi.org/10.1103/PhysRevB.95.094424} {\bibfield  {journal}
  {\bibinfo  {journal} {Phys. Rev. B}\ }\textbf {\bibinfo {volume} {95}},\
  \bibinfo {pages} {094424} (\bibinfo {year} {2017})}\BibitemShut {NoStop}%
\bibitem [{\citenamefont {Zhong}\ \emph {et~al.}(2020)\citenamefont {Zhong},
  \citenamefont {Gao}, \citenamefont {Ong},\ and\ \citenamefont
  {Cava}}]{zhong2020weak}%
  \BibitemOpen
  \bibfield  {author} {\bibinfo {author} {\bibfnamefont {R.}~\bibnamefont
  {Zhong}}, \bibinfo {author} {\bibfnamefont {T.}~\bibnamefont {Gao}}, \bibinfo
  {author} {\bibfnamefont {N.~P.}\ \bibnamefont {Ong}},\ and\ \bibinfo {author}
  {\bibfnamefont {R.~J.}\ \bibnamefont {Cava}},\ }\bibfield  {title} {\bibinfo
  {title} {Weak-field induced nonmagnetic state in a {C}o-based honeycomb},\
  }\href {https://doi.org/10.1126/sciadv.aay6953} {\bibfield  {journal}
  {\bibinfo  {journal} {Science Advances}\ }\textbf {\bibinfo {volume} {6}},\
  \bibinfo {pages} {eaay6953} (\bibinfo {year} {2020})}\BibitemShut {NoStop}%
\bibitem [{\citenamefont {Zhang}\ \emph {et~al.}(2023)\citenamefont {Zhang},
  \citenamefont {Xu}, \citenamefont {Halloran}, \citenamefont {Zhong},
  \citenamefont {Broholm}, \citenamefont {Cava}, \citenamefont {Drichko},\ and\
  \citenamefont {Armitage}}]{Zhang2023}%
  \BibitemOpen
  \bibfield  {author} {\bibinfo {author} {\bibfnamefont {X.}~\bibnamefont
  {Zhang}}, \bibinfo {author} {\bibfnamefont {Y.}~\bibnamefont {Xu}}, \bibinfo
  {author} {\bibfnamefont {T.}~\bibnamefont {Halloran}}, \bibinfo {author}
  {\bibfnamefont {R.}~\bibnamefont {Zhong}}, \bibinfo {author} {\bibfnamefont
  {C.}~\bibnamefont {Broholm}}, \bibinfo {author} {\bibfnamefont {R.~J.}\
  \bibnamefont {Cava}}, \bibinfo {author} {\bibfnamefont {N.}~\bibnamefont
  {Drichko}},\ and\ \bibinfo {author} {\bibfnamefont {N.~P.}\ \bibnamefont
  {Armitage}},\ }\bibfield  {title} {\bibinfo {title} {A magnetic continuum in
  the cobalt-based honeycomb magnet {B}a{C}o$_2$({A}s{O}$_4$)$_2$},\ }\href
  {https://doi.org/10.1038/s41563-022-01403-1} {\bibfield  {journal} {\bibinfo
  {journal} {Nat. Mater.}\ }\textbf {\bibinfo {volume} {22}},\ \bibinfo {pages}
  {58} (\bibinfo {year} {2023})}\BibitemShut {NoStop}%
\bibitem [{\citenamefont {Halloran}\ \emph {et~al.}(2023)\citenamefont
  {Halloran}, \citenamefont {Desrochers}, \citenamefont {Zhang}, \citenamefont
  {Chen}, \citenamefont {Chern}, \citenamefont {Xu}, \citenamefont {Winn},
  \citenamefont {Graves-Brook}, \citenamefont {Stone}, \citenamefont
  {Kolesnikov}, \citenamefont {Qiu}, \citenamefont {Zhong}, \citenamefont
  {Cava}, \citenamefont {Kim},\ and\ \citenamefont {Broholm}}]{Halloran23}%
  \BibitemOpen
  \bibfield  {author} {\bibinfo {author} {\bibfnamefont {T.}~\bibnamefont
  {Halloran}}, \bibinfo {author} {\bibfnamefont {F.}~\bibnamefont
  {Desrochers}}, \bibinfo {author} {\bibfnamefont {E.~Z.}\ \bibnamefont
  {Zhang}}, \bibinfo {author} {\bibfnamefont {T.}~\bibnamefont {Chen}},
  \bibinfo {author} {\bibfnamefont {L.~E.}\ \bibnamefont {Chern}}, \bibinfo
  {author} {\bibfnamefont {Z.}~\bibnamefont {Xu}}, \bibinfo {author}
  {\bibfnamefont {B.}~\bibnamefont {Winn}}, \bibinfo {author} {\bibfnamefont
  {M.}~\bibnamefont {Graves-Brook}}, \bibinfo {author} {\bibfnamefont {M.~B.}\
  \bibnamefont {Stone}}, \bibinfo {author} {\bibfnamefont {A.~I.}\ \bibnamefont
  {Kolesnikov}}, \bibinfo {author} {\bibfnamefont {Y.}~\bibnamefont {Qiu}},
  \bibinfo {author} {\bibfnamefont {R.}~\bibnamefont {Zhong}}, \bibinfo
  {author} {\bibfnamefont {R.}~\bibnamefont {Cava}}, \bibinfo {author}
  {\bibfnamefont {Y.~B.}\ \bibnamefont {Kim}},\ and\ \bibinfo {author}
  {\bibfnamefont {C.}~\bibnamefont {Broholm}},\ }\bibfield  {title} {\bibinfo
  {title} {Geometrical frustration versus {K}itaev interactions in
  {B}a{C}o$_2$({A}s{O}$_4$)$_2$},\ }\href
  {https://doi.org/10.1073/pnas.2215509119} {\bibfield  {journal} {\bibinfo
  {journal} {Proc. Natl. Acad. Sci.}\ }\textbf {\bibinfo {volume} {120}},\
  \bibinfo {pages} {e2215509119} (\bibinfo {year} {2023})}\BibitemShut
  {NoStop}%
\bibitem [{\citenamefont {Xiang}\ \emph {et~al.}(2023)\citenamefont {Xiang},
  \citenamefont {Dhakal}, \citenamefont {Ozerov}, \citenamefont {Jiang},
  \citenamefont {Mou}, \citenamefont {Ozarowski}, \citenamefont {Huang},
  \citenamefont {Zhou}, \citenamefont {Fang}, \citenamefont {Winter},
  \citenamefont {Jiang},\ and\ \citenamefont {Smirnov}}]{Xiang2023Disorder}%
  \BibitemOpen
  \bibfield  {author} {\bibinfo {author} {\bibfnamefont {L.}~\bibnamefont
  {Xiang}}, \bibinfo {author} {\bibfnamefont {R.}~\bibnamefont {Dhakal}},
  \bibinfo {author} {\bibfnamefont {M.}~\bibnamefont {Ozerov}}, \bibinfo
  {author} {\bibfnamefont {Y.}~\bibnamefont {Jiang}}, \bibinfo {author}
  {\bibfnamefont {B.~S.}\ \bibnamefont {Mou}}, \bibinfo {author} {\bibfnamefont
  {A.}~\bibnamefont {Ozarowski}}, \bibinfo {author} {\bibfnamefont
  {Q.}~\bibnamefont {Huang}}, \bibinfo {author} {\bibfnamefont
  {H.}~\bibnamefont {Zhou}}, \bibinfo {author} {\bibfnamefont {J.}~\bibnamefont
  {Fang}}, \bibinfo {author} {\bibfnamefont {S.~M.}\ \bibnamefont {Winter}},
  \bibinfo {author} {\bibfnamefont {Z.}~\bibnamefont {Jiang}},\ and\ \bibinfo
  {author} {\bibfnamefont {D.}~\bibnamefont {Smirnov}},\ }\bibfield  {title}
  {\bibinfo {title} {Disorder-enriched magnetic excitations in a
  {H}eisenberg-{K}itaev quantum magnet {N}a$_{2}${C}o$_{2}${T}e{O}$_{6}$},\
  }\href {https://doi.org/10.1103/PhysRevLett.131.076701} {\bibfield  {journal}
  {\bibinfo  {journal} {Phys. Rev. Lett.}\ }\textbf {\bibinfo {volume} {131}},\
  \bibinfo {pages} {076701} (\bibinfo {year} {2023})}\BibitemShut {NoStop}%
\bibitem [{\citenamefont {Pilch}\ \emph {et~al.}(2023)\citenamefont {Pilch},
  \citenamefont {Peedu}, \citenamefont {Bera}, \citenamefont {Yusuf},
  \citenamefont {Nagel}, \citenamefont {R\~{o}\~om},\ and\ \citenamefont
  {Wang}}]{Pilch2023_Field}%
  \BibitemOpen
  \bibfield  {author} {\bibinfo {author} {\bibfnamefont {P.}~\bibnamefont
  {Pilch}}, \bibinfo {author} {\bibfnamefont {L.}~\bibnamefont {Peedu}},
  \bibinfo {author} {\bibfnamefont {A.~K.}\ \bibnamefont {Bera}}, \bibinfo
  {author} {\bibfnamefont {S.~M.}\ \bibnamefont {Yusuf}}, \bibinfo {author}
  {\bibfnamefont {U.}~\bibnamefont {Nagel}}, \bibinfo {author} {\bibfnamefont
  {T.}~\bibnamefont {R\~{o}\~om}},\ and\ \bibinfo {author} {\bibfnamefont
  {Z.}~\bibnamefont {Wang}},\ }\bibfield  {title} {\bibinfo {title} {Field- and
  polarization-dependent quantum spin dynamics in the honeycomb magnet
  {N}a$_{2}${C}o$_{2}${T}e{O}$_{6}$: {M}agnetic excitations and continuum},\
  }\href {https://doi.org/10.1103/PhysRevB.108.L140406} {\bibfield  {journal}
  {\bibinfo  {journal} {Phys. Rev. B}\ }\textbf {\bibinfo {volume} {108}},\
  \bibinfo {pages} {L140406} (\bibinfo {year} {2023})}\BibitemShut {NoStop}%
\bibitem [{\citenamefont {Wang}\ \emph {et~al.}(2017)\citenamefont {Wang},
  \citenamefont {Reschke}, \citenamefont {H\"uvonen}, \citenamefont {Do},
  \citenamefont {Choi}, \citenamefont {Gensch}, \citenamefont {Nagel},
  \citenamefont {R\~{o}\~om},\ and\ \citenamefont
  {Loidl}}]{Wang2017alphaRuCl_FTIR}%
  \BibitemOpen
  \bibfield  {author} {\bibinfo {author} {\bibfnamefont {Z.}~\bibnamefont
  {Wang}}, \bibinfo {author} {\bibfnamefont {S.}~\bibnamefont {Reschke}},
  \bibinfo {author} {\bibfnamefont {D.}~\bibnamefont {H\"uvonen}}, \bibinfo
  {author} {\bibfnamefont {S.-H.}\ \bibnamefont {Do}}, \bibinfo {author}
  {\bibfnamefont {K.-Y.}\ \bibnamefont {Choi}}, \bibinfo {author}
  {\bibfnamefont {M.}~\bibnamefont {Gensch}}, \bibinfo {author} {\bibfnamefont
  {U.}~\bibnamefont {Nagel}}, \bibinfo {author} {\bibfnamefont
  {T.}~\bibnamefont {R\~{o}\~om}},\ and\ \bibinfo {author} {\bibfnamefont
  {A.}~\bibnamefont {Loidl}},\ }\bibfield  {title} {\bibinfo {title} {Magnetic
  excitations and continuum of a possibly field-induced quantum spin liquid in
  $\ensuremath{\alpha}\text{\ensuremath{-}}${R}u{C}l$_{3}$},\ }\href
  {https://doi.org/10.1103/PhysRevLett.119.227202} {\bibfield  {journal}
  {\bibinfo  {journal} {Phys. Rev. Lett.}\ }\textbf {\bibinfo {volume} {119}},\
  \bibinfo {pages} {227202} (\bibinfo {year} {2017})}\BibitemShut {NoStop}%
\bibitem [{\citenamefont {Banerjee}\ \emph {et~al.}(2018)\citenamefont
  {Banerjee}, \citenamefont {Lampen-Kelley}, \citenamefont {Knolle},
  \citenamefont {Balz}, \citenamefont {Aczel}, \citenamefont {Winn},
  \citenamefont {Liu}, \citenamefont {Pajerowski}, \citenamefont {Yan},
  \citenamefont {Bridges}, \citenamefont {Savici}, \citenamefont {Chakoumakos},
  \citenamefont {Lumsden}, \citenamefont {Tennant}, \citenamefont {Moessner},
  \citenamefont {Mandrus},\ and\ \citenamefont {Nagler}}]{Banerjee2018}%
  \BibitemOpen
  \bibfield  {author} {\bibinfo {author} {\bibfnamefont {A.}~\bibnamefont
  {Banerjee}}, \bibinfo {author} {\bibfnamefont {P.}~\bibnamefont
  {Lampen-Kelley}}, \bibinfo {author} {\bibfnamefont {J.}~\bibnamefont
  {Knolle}}, \bibinfo {author} {\bibfnamefont {C.}~\bibnamefont {Balz}},
  \bibinfo {author} {\bibfnamefont {A.~A.}\ \bibnamefont {Aczel}}, \bibinfo
  {author} {\bibfnamefont {B.}~\bibnamefont {Winn}}, \bibinfo {author}
  {\bibfnamefont {Y.}~\bibnamefont {Liu}}, \bibinfo {author} {\bibfnamefont
  {D.}~\bibnamefont {Pajerowski}}, \bibinfo {author} {\bibfnamefont
  {J.}~\bibnamefont {Yan}}, \bibinfo {author} {\bibfnamefont {C.~A.}\
  \bibnamefont {Bridges}}, \bibinfo {author} {\bibfnamefont {A.~T.}\
  \bibnamefont {Savici}}, \bibinfo {author} {\bibfnamefont {B.~C.}\
  \bibnamefont {Chakoumakos}}, \bibinfo {author} {\bibfnamefont {M.~D.}\
  \bibnamefont {Lumsden}}, \bibinfo {author} {\bibfnamefont {D.~A.}\
  \bibnamefont {Tennant}}, \bibinfo {author} {\bibfnamefont {R.}~\bibnamefont
  {Moessner}}, \bibinfo {author} {\bibfnamefont {D.~G.}\ \bibnamefont
  {Mandrus}},\ and\ \bibinfo {author} {\bibfnamefont {S.~E.}\ \bibnamefont
  {Nagler}},\ }\bibfield  {title} {\bibinfo {title} {Excitations in the
  field-induced quantum spin liquid state of $\alpha$-{R}u{C}l$_3$},\ }\href
  {https://doi.org/10.1038/s41535-018-0079-2} {\bibfield  {journal} {\bibinfo
  {journal} {npj Quantum Mater.}\ }\textbf {\bibinfo {volume} {3}},\ \bibinfo
  {pages} {8} (\bibinfo {year} {2018})}\BibitemShut {NoStop}%
\bibitem [{\citenamefont {Xu}\ \emph {et~al.}(2023)\citenamefont {Xu},
  \citenamefont {Wu}, \citenamefont {Zhu},\ and\ \citenamefont
  {Cava}}]{Xu2023KCTO}%
  \BibitemOpen
  \bibfield  {author} {\bibinfo {author} {\bibfnamefont {X.}~\bibnamefont
  {Xu}}, \bibinfo {author} {\bibfnamefont {L.}~\bibnamefont {Wu}}, \bibinfo
  {author} {\bibfnamefont {Y.}~\bibnamefont {Zhu}},\ and\ \bibinfo {author}
  {\bibfnamefont {R.~J.}\ \bibnamefont {Cava}},\ }\bibfield  {title} {\bibinfo
  {title} {{K}$_{2}${C}o$_{2}${T}e{O}$_{6}$: A layered magnet with a
  ${S}=\frac{1}{2}$ {C}o$^{2+}$ honeycomb lattice},\ }\href
  {https://doi.org/10.1103/PhysRevB.108.174432} {\bibfield  {journal} {\bibinfo
   {journal} {Phys. Rev. B}\ }\textbf {\bibinfo {volume} {108}},\ \bibinfo
  {pages} {174432} (\bibinfo {year} {2023})}\BibitemShut {NoStop}%
\bibitem [{\citenamefont {Zhang}\ \emph {et~al.}(2020)\citenamefont {Zhang},
  \citenamefont {Amelin}, \citenamefont {Wang}, \citenamefont {Zou},
  \citenamefont {Yang}, \citenamefont {Nagel}, \citenamefont {R\~{o}\~om},
  \citenamefont {Dey}, \citenamefont {Nugroho}, \citenamefont {Lorenz},
  \citenamefont {Wu},\ and\ \citenamefont {Wang}}]{Zhang20}%
  \BibitemOpen
  \bibfield  {author} {\bibinfo {author} {\bibfnamefont {Z.}~\bibnamefont
  {Zhang}}, \bibinfo {author} {\bibfnamefont {K.}~\bibnamefont {Amelin}},
  \bibinfo {author} {\bibfnamefont {X.}~\bibnamefont {Wang}}, \bibinfo {author}
  {\bibfnamefont {H.}~\bibnamefont {Zou}}, \bibinfo {author} {\bibfnamefont
  {J.}~\bibnamefont {Yang}}, \bibinfo {author} {\bibfnamefont {U.}~\bibnamefont
  {Nagel}}, \bibinfo {author} {\bibfnamefont {T.}~\bibnamefont {R\~{o}\~om}},
  \bibinfo {author} {\bibfnamefont {T.}~\bibnamefont {Dey}}, \bibinfo {author}
  {\bibfnamefont {A.~A.}\ \bibnamefont {Nugroho}}, \bibinfo {author}
  {\bibfnamefont {T.}~\bibnamefont {Lorenz}}, \bibinfo {author} {\bibfnamefont
  {J.}~\bibnamefont {Wu}},\ and\ \bibinfo {author} {\bibfnamefont
  {Z.}~\bibnamefont {Wang}},\ }\bibfield  {title} {\bibinfo {title}
  {Observation of ${E}_{8}$ particles in an {I}sing chain antiferromagnet},\
  }\href {https://doi.org/10.1103/PhysRevB.101.220411} {\bibfield  {journal}
  {\bibinfo  {journal} {Phys. Rev. B}\ }\textbf {\bibinfo {volume} {101}},\
  \bibinfo {pages} {220411} (\bibinfo {year} {2020})}\BibitemShut {NoStop}%
\bibitem [{\citenamefont {Amelin}\ \emph {et~al.}(2020)\citenamefont {Amelin},
  \citenamefont {Engelmayer}, \citenamefont {Viirok}, \citenamefont {Nagel},
  \citenamefont {R\~{o}\~om}, \citenamefont {Lorenz},\ and\ \citenamefont
  {Wang}}]{Amelin20}%
  \BibitemOpen
  \bibfield  {author} {\bibinfo {author} {\bibfnamefont {K.}~\bibnamefont
  {Amelin}}, \bibinfo {author} {\bibfnamefont {J.}~\bibnamefont {Engelmayer}},
  \bibinfo {author} {\bibfnamefont {J.}~\bibnamefont {Viirok}}, \bibinfo
  {author} {\bibfnamefont {U.}~\bibnamefont {Nagel}}, \bibinfo {author}
  {\bibfnamefont {T.}~\bibnamefont {R\~{o}\~om}}, \bibinfo {author}
  {\bibfnamefont {T.}~\bibnamefont {Lorenz}},\ and\ \bibinfo {author}
  {\bibfnamefont {Z.}~\bibnamefont {Wang}},\ }\bibfield  {title} {\bibinfo
  {title} {Experimental observation of quantum many-body excitations of
  ${E}_{8}$ symmetry in the {I}sing chain ferromagnet
  {C}o{N}b$_{2}${O}$_{6}$},\ }\href
  {https://doi.org/10.1103/PhysRevB.102.104431} {\bibfield  {journal} {\bibinfo
   {journal} {Phys. Rev. B}\ }\textbf {\bibinfo {volume} {102}},\ \bibinfo
  {pages} {104431} (\bibinfo {year} {2020})}\BibitemShut {NoStop}%
\bibitem [{\citenamefont {Amelin}\ \emph {et~al.}(2022)\citenamefont {Amelin},
  \citenamefont {Viirok}, \citenamefont {Nagel}, \citenamefont {R\~{o}\~om},
  \citenamefont {Engelmayer}, \citenamefont {Dey}, \citenamefont {Nugroho},
  \citenamefont {Lorenz},\ and\ \citenamefont {Wang}}]{Amelin22}%
  \BibitemOpen
  \bibfield  {author} {\bibinfo {author} {\bibfnamefont {K.}~\bibnamefont
  {Amelin}}, \bibinfo {author} {\bibfnamefont {J.}~\bibnamefont {Viirok}},
  \bibinfo {author} {\bibfnamefont {U.}~\bibnamefont {Nagel}}, \bibinfo
  {author} {\bibfnamefont {T.}~\bibnamefont {R\~{o}\~om}}, \bibinfo {author}
  {\bibfnamefont {J.}~\bibnamefont {Engelmayer}}, \bibinfo {author}
  {\bibfnamefont {T.}~\bibnamefont {Dey}}, \bibinfo {author} {\bibfnamefont
  {A.~A.}\ \bibnamefont {Nugroho}}, \bibinfo {author} {\bibfnamefont
  {T.}~\bibnamefont {Lorenz}},\ and\ \bibinfo {author} {\bibfnamefont
  {Z.}~\bibnamefont {Wang}},\ }\bibfield  {title} {\bibinfo {title} {Quantum
  spin dynamics of quasi-one-dimensional {H}eisenberg-{I}sing magnets in a
  transverse field: confined spinons, ${E}_8$ spectrum, and quantum phase
  transitions},\ }\href {https://doi.org/10.1088/1751-8121/aca6b8} {\bibfield
  {journal} {\bibinfo  {journal} {J. Phys. A: Math. Theor.}\ }\textbf {\bibinfo
  {volume} {55}},\ \bibinfo {pages} {484005} (\bibinfo {year}
  {2022})}\BibitemShut {NoStop}%
\bibitem [{\citenamefont {Pilch}\ \emph {et~al.}(2025)\citenamefont {Pilch},
  \citenamefont {Amelin}, \citenamefont {Schmiedinghoff}, \citenamefont
  {Reinold}, \citenamefont {Zhu}, \citenamefont {Povarov}, \citenamefont
  {Zvyagin}, \citenamefont {Engelkamp}, \citenamefont {Lan}, \citenamefont
  {Shu}, \citenamefont {Chou}, \citenamefont {Nagel}, \citenamefont
  {R\~{o}\~om}, \citenamefont {Uhrig}, \citenamefont {Fauseweh},\ and\
  \citenamefont {Wang}}]{Pilch25}%
  \BibitemOpen
  \bibfield  {author} {\bibinfo {author} {\bibfnamefont {P.}~\bibnamefont
  {Pilch}}, \bibinfo {author} {\bibfnamefont {K.}~\bibnamefont {Amelin}},
  \bibinfo {author} {\bibfnamefont {G.}~\bibnamefont {Schmiedinghoff}},
  \bibinfo {author} {\bibfnamefont {A.}~\bibnamefont {Reinold}}, \bibinfo
  {author} {\bibfnamefont {C.}~\bibnamefont {Zhu}}, \bibinfo {author}
  {\bibfnamefont {K.~Y.}\ \bibnamefont {Povarov}}, \bibinfo {author}
  {\bibfnamefont {S.}~\bibnamefont {Zvyagin}}, \bibinfo {author} {\bibfnamefont
  {H.}~\bibnamefont {Engelkamp}}, \bibinfo {author} {\bibfnamefont {Y.-P.}\
  \bibnamefont {Lan}}, \bibinfo {author} {\bibfnamefont {G.-J.}\ \bibnamefont
  {Shu}}, \bibinfo {author} {\bibfnamefont {F.~C.}\ \bibnamefont {Chou}},
  \bibinfo {author} {\bibfnamefont {U.}~\bibnamefont {Nagel}}, \bibinfo
  {author} {\bibfnamefont {T.}~\bibnamefont {R\~{o}\~om}}, \bibinfo {author}
  {\bibfnamefont {G.~S.}\ \bibnamefont {Uhrig}}, \bibinfo {author}
  {\bibfnamefont {B.}~\bibnamefont {Fauseweh}},\ and\ \bibinfo {author}
  {\bibfnamefont {Z.}~\bibnamefont {Wang}},\ }\bibfield  {title} {\bibinfo
  {title} {Low-energy spin excitations in field-induced phases of the
  spin-ladder antiferromagnet {B}i{C}u$_{2}${PO}$_{6}$},\ }\href
  {https://doi.org/10.1103/PhysRevB.111.024423} {\bibfield  {journal} {\bibinfo
   {journal} {Phys. Rev. B}\ }\textbf {\bibinfo {volume} {111}},\ \bibinfo
  {pages} {024423} (\bibinfo {year} {2025})}\BibitemShut {NoStop}%
\bibitem [{\citenamefont {Lin}\ \emph {et~al.}(2021)\citenamefont {Lin},
  \citenamefont {Jeong}, \citenamefont {Kim}, \citenamefont {Wang},
  \citenamefont {Huang}, \citenamefont {Masuda}, \citenamefont {Asai},
  \citenamefont {Itoh}, \citenamefont {G{\"u}nther}, \citenamefont {Russina}
  \emph {et~al.}}]{lin2021field}%
  \BibitemOpen
  \bibfield  {author} {\bibinfo {author} {\bibfnamefont {G.}~\bibnamefont
  {Lin}}, \bibinfo {author} {\bibfnamefont {J.}~\bibnamefont {Jeong}}, \bibinfo
  {author} {\bibfnamefont {C.}~\bibnamefont {Kim}}, \bibinfo {author}
  {\bibfnamefont {Y.}~\bibnamefont {Wang}}, \bibinfo {author} {\bibfnamefont
  {Q.}~\bibnamefont {Huang}}, \bibinfo {author} {\bibfnamefont
  {T.}~\bibnamefont {Masuda}}, \bibinfo {author} {\bibfnamefont
  {S.}~\bibnamefont {Asai}}, \bibinfo {author} {\bibfnamefont {S.}~\bibnamefont
  {Itoh}}, \bibinfo {author} {\bibfnamefont {G.}~\bibnamefont {G{\"u}nther}},
  \bibinfo {author} {\bibfnamefont {M.}~\bibnamefont {Russina}}, \emph
  {et~al.},\ }\bibfield  {title} {\bibinfo {title} {Field-induced quantum spin
  disordered state in spin-1/2 honeycomb magnet {N}a$_2${C}o$_2${T}e{O}$_6$},\
  }\href {https://doi.org/10.1038/s41467-021-25567-7} {\bibfield  {journal}
  {\bibinfo  {journal} {Nat. Commun.}\ }\textbf {\bibinfo {volume} {12}},\
  \bibinfo {pages} {5559} (\bibinfo {year} {2021})}\BibitemShut {NoStop}%
\bibitem [{\citenamefont {Greger}\ \emph {et~al.}(2013)\citenamefont {Greger},
  \citenamefont {Kollar},\ and\ \citenamefont {Vollhardt}}]{Greger13}%
  \BibitemOpen
  \bibfield  {author} {\bibinfo {author} {\bibfnamefont {M.}~\bibnamefont
  {Greger}}, \bibinfo {author} {\bibfnamefont {M.}~\bibnamefont {Kollar}},\
  and\ \bibinfo {author} {\bibfnamefont {D.}~\bibnamefont {Vollhardt}},\
  }\bibfield  {title} {\bibinfo {title} {Isosbestic points: How a narrow
  crossing region of curves determines their leading parameter dependence},\
  }\href {https://doi.org/10.1103/PhysRevB.87.195140} {\bibfield  {journal}
  {\bibinfo  {journal} {Phys. Rev. B}\ }\textbf {\bibinfo {volume} {87}},\
  \bibinfo {pages} {195140} (\bibinfo {year} {2013})}\BibitemShut {NoStop}%
\bibitem [{\citenamefont {Vollhardt}(1997)}]{Vollhardt97}%
  \BibitemOpen
  \bibfield  {author} {\bibinfo {author} {\bibfnamefont {D.}~\bibnamefont
  {Vollhardt}},\ }\bibfield  {title} {\bibinfo {title} {Characteristic crossing
  points in specific heat curves of correlated systems},\ }\href
  {https://doi.org/10.1103/PhysRevLett.78.1307} {\bibfield  {journal} {\bibinfo
   {journal} {Phys. Rev. Lett.}\ }\textbf {\bibinfo {volume} {78}},\ \bibinfo
  {pages} {1307} (\bibinfo {year} {1997})}\BibitemShut {NoStop}%
\bibitem [{\citenamefont {Wang}\ \emph {et~al.}(2014)\citenamefont {Wang},
  \citenamefont {Schmidt}, \citenamefont {Fischer}, \citenamefont {Tsurkan},
  \citenamefont {Greger}, \citenamefont {Vollhardt}, \citenamefont {Loidl},\
  and\ \citenamefont {Deisenhofer}}]{Wang2014}%
  \BibitemOpen
  \bibfield  {author} {\bibinfo {author} {\bibfnamefont {Z.}~\bibnamefont
  {Wang}}, \bibinfo {author} {\bibfnamefont {M.}~\bibnamefont {Schmidt}},
  \bibinfo {author} {\bibfnamefont {J.}~\bibnamefont {Fischer}}, \bibinfo
  {author} {\bibfnamefont {V.}~\bibnamefont {Tsurkan}}, \bibinfo {author}
  {\bibfnamefont {M.}~\bibnamefont {Greger}}, \bibinfo {author} {\bibfnamefont
  {D.}~\bibnamefont {Vollhardt}}, \bibinfo {author} {\bibfnamefont
  {A.}~\bibnamefont {Loidl}},\ and\ \bibinfo {author} {\bibfnamefont
  {J.}~\bibnamefont {Deisenhofer}},\ }\bibfield  {title} {\bibinfo {title}
  {Orbital-selective metal–insulator transition and gap formation above
  {T}$_c$ in superconducting {R}b$_{1-x}${F}e$_{2-y}${S}e$_2$},\ }\href
  {https://doi.org/10.1038/ncomms4202} {\bibfield  {journal} {\bibinfo
  {journal} {Nat. Commun.}\ }\textbf {\bibinfo {volume} {5}},\ \bibinfo {pages}
  {3202} (\bibinfo {year} {2014})}\BibitemShut {NoStop}%
\bibitem [{\citenamefont {Kovalev}\ \emph {et~al.}(2021)\citenamefont
  {Kovalev}, \citenamefont {Dong}, \citenamefont {Shi}, \citenamefont
  {Reinhoffer}, \citenamefont {Xu}, \citenamefont {Wang}, \citenamefont {Wang},
  \citenamefont {Gan}, \citenamefont {Germanskiy}, \citenamefont {Deinert},
  \citenamefont {Ilyakov}, \citenamefont {van Loosdrecht}, \citenamefont {Wu},
  \citenamefont {Wang}, \citenamefont {Demsar},\ and\ \citenamefont
  {Wang}}]{Kovalev21}%
  \BibitemOpen
  \bibfield  {author} {\bibinfo {author} {\bibfnamefont {S.}~\bibnamefont
  {Kovalev}}, \bibinfo {author} {\bibfnamefont {T.}~\bibnamefont {Dong}},
  \bibinfo {author} {\bibfnamefont {L.-Y.}\ \bibnamefont {Shi}}, \bibinfo
  {author} {\bibfnamefont {C.}~\bibnamefont {Reinhoffer}}, \bibinfo {author}
  {\bibfnamefont {T.-Q.}\ \bibnamefont {Xu}}, \bibinfo {author} {\bibfnamefont
  {H.-Z.}\ \bibnamefont {Wang}}, \bibinfo {author} {\bibfnamefont
  {Y.}~\bibnamefont {Wang}}, \bibinfo {author} {\bibfnamefont {Z.-Z.}\
  \bibnamefont {Gan}}, \bibinfo {author} {\bibfnamefont {S.}~\bibnamefont
  {Germanskiy}}, \bibinfo {author} {\bibfnamefont {J.-C.}\ \bibnamefont
  {Deinert}}, \bibinfo {author} {\bibfnamefont {I.}~\bibnamefont {Ilyakov}},
  \bibinfo {author} {\bibfnamefont {P.~H.~M.}\ \bibnamefont {van Loosdrecht}},
  \bibinfo {author} {\bibfnamefont {D.}~\bibnamefont {Wu}}, \bibinfo {author}
  {\bibfnamefont {N.-L.}\ \bibnamefont {Wang}}, \bibinfo {author}
  {\bibfnamefont {J.}~\bibnamefont {Demsar}},\ and\ \bibinfo {author}
  {\bibfnamefont {Z.}~\bibnamefont {Wang}},\ }\bibfield  {title} {\bibinfo
  {title} {Band-selective third-harmonic generation in superconducting
  {M}g{B}$_{2}$: Possible evidence for the {H}iggs amplitude mode in the dirty
  limit},\ }\href {https://doi.org/10.1103/PhysRevB.104.L140505} {\bibfield
  {journal} {\bibinfo  {journal} {Phys. Rev. B}\ }\textbf {\bibinfo {volume}
  {104}},\ \bibinfo {pages} {L140505} (\bibinfo {year} {2021})}\BibitemShut
  {NoStop}%
\end{thebibliography}%

\end{document}